\documentclass[twocolumn,english,prl,superscriptaddress,showpacs]{revtex4}
\usepackage{times}
\usepackage[T1]{fontenc}
\usepackage[latin1]{inputenc}
\usepackage{amsmath}
\usepackage{graphicx}
\usepackage{amssymb}

\makeatletter

\usepackage{babel}
\makeatother
\begin{document}

\title{Switchable lasing in coupled multimode microcavities}

\author{Sergei V. Zhukovsky}

\email{sergei@physik.uni-bonn.de}

\affiliation{Physikalisches Institut, Universit\"at Bonn, Nussallee 12, D-53115
Bonn, Germany}

\author{Dmitry N. Chigrin}


\affiliation{Physikalisches Institut, Universit\"at Bonn, Nussallee 12, D-53115
Bonn, Germany}

\author{Andrei V. Lavrinenko}


\affiliation{COM-DTU, Department of Communications, Optics and Materials, NanoDTU,
Technical University of Denmark, Building 345V, DK-2800 Kgs.~Lyngby,
Denmark}

\author{Johann Kroha}


\affiliation{Physikalisches Institut, Universit\"at Bonn, Nussallee 12, D-53115
Bonn, Germany}

\begin{abstract} 
We propose the new concept of a switchable multimode microlaser. 
As a generic, realistic model of a multimode microresonator 
a system of two coupled defects in a two-dimensional photonic crystal is 
considered. We demonstrate theoretically
that lasing of the cavity into one selected resonator mode can be caused 
by injecting an appropriate optical pulse at the onset of laser action
(injection seeding). Temporal mode-to-mode switching
by re-seeding the cavity after a short cool-down period is demonstrated
by direct numerical solution. A qualitative analytical explanation
of the mode switching in terms of the laser bistability is presented.
\end{abstract}


\maketitle
Microlasers with cavity sizes comparable to the radiation wavelength
$\lambda$ are very promising from both fundamental and application points of
view \cite{VahalaReview} for use as integrated coherent light sources. 
Making microlasers capable of multiple-wavelength
emission contributes even more towards miniaturization of optical
components, and also provides an additional degree of freedom in
light control.  The common approach towards microlaser tunability is in
essence modification of the optical properties of a single-mode
cavity by thermal \cite{tune2therm}, micromechanical \cite{tune3mech}
or electrooptical means \cite{tune4}.


In this Letter we propose, in contrast, the new concept of a 
\emph{switchable microlaser}, comprised of a multimode laser microresonator,
where lasing can be switched on demand to any of its eigenmodes.
While the broad gain profile of semiconductor and dye lasers provides
little discrimination between neighboring modes and, thus, leads 
to mode hopping and multistable mode dynamics, 
we show that a definite resonator mode can be selected for lasing 
by \emph{injection seeding} \cite{Siegman,seeding}, i.e. by injecting an
appropriate pulse before and during the onset of lasing, such that the 
stimulated emission builds up in a chosen mode from this seeding field rather 
than from the random noise present in the system due to 
quantum fluctuations and spontaneous emission. 
We investigate the time needed for switching between different lasing 
modes and analyze how this process is influenced by
noise in competition with the seeding signal.  

\begin{figure}[b]
\includegraphics[clip,width=0.9\columnwidth]{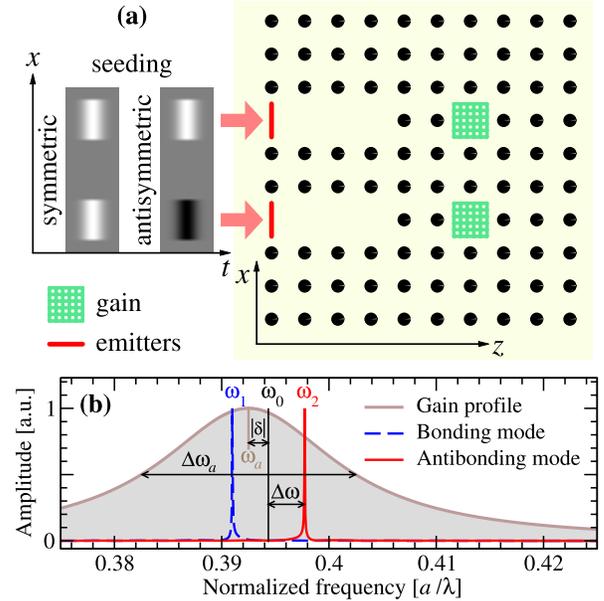}
\caption{(Color online) 
a) Schematic illustration of the resonant system under
study. Two coupled defects in a periodic 2D PhC lattice contain a
gain medium. Two emitters shown are used to produce the seeding signal
delivered to the defect sites by two waveguides. The spatial field
distribution of the seeding signal is schematically depicted as a 
gray scale on the left.
b) Frequency parameters of the two-mode laser, see text. 
\label{fig:setup}}
\end{figure}

We generalize the semiclassical 
multimode laser model of Ref.~\cite{Hodges} to study the dynamics
of individual modes in a two-mode laser, where an external field is present
in the cavity due to the injection seeding. 
Semiclassical theories of optical amplifiers 
\cite{Siegman} 
have been successfully employed to study the mode dynamics in multimode
lasers \cite{Hodges,JohnBusch}, 
especially in the context of laser instabilities \cite{Abraham}.
An example of a two-mode
microcavity, which we have chosen to study numerically, 
is depicted in Fig.~\ref{fig:setup} a).
The microcavity is based on two coupled defects in a two-dimensional
(2D) photonic crystal (PhC) made of dielectric nanopillars \cite{deSterke1}.
Pillar PhCs are practically feasible by state-of-the-art
fabrication technology \cite{pillar-lastexp}.

The system of two identical coupled single-mode cavities supports
two modes, namely the bonding (symmetric) and the antibonding
(antisymmetric) mode, characterized by spatial field
distributions $u_{1,2}(\mathbf{r})$ and
frequencies $\omega_{1,2}=\omega_{0}\mp\Delta\omega$,
respectively. Here $\Delta\omega$ is the mode detuning from the frequency
of the single-cavity resonance, $\omega_{0}$. For weak mode overlap the spatial
intensity profiles of two modes nearly coincide, 
$\left|u_{1}(r)\right|^{2} \approx \left|u_{2}(r)\right|^{2}$.
We assume that the cavities contain a laser medium with the gain centered at
frequency~$\omega_{a}=\omega_{0}+\delta$, homogeneously broadened
to have a line width of~$\Delta\omega_{a}>\Delta\omega$,
where $\delta$ is the detuning
of the gain profile from the cavity frequency $\omega_{0}$. 
See Fig.~\ref{fig:setup} b) for the definition of the frequency parameters. 
For the numerical calculations below we take a quadratic array of 
$12 \times 12$ nanopillars with dielectric constant $\epsilon=9$  in
air ($\epsilon =1$). The lattice period is $a=500$~nm and the pillar radius
$r=0.2 a$. Then both defect modes have frequencies near 
$a/\lambda\simeq0.395$ [Fig.~\ref{fig:setup} b)].

To obtain an initial understanding of the seeding-induced mode switching,
the semiclassical Maxwell-Bloch equations \cite{Siegman},
may be simplyfied using the rotating-wave and the
slowly varying envelope approximations. 
All the spatial dependencies of the electric field and atomic polarization
can then be represented 
in the basis of the two cavity modes, such that $E(\mathbf{r},t)=E_{1}(t)u_{1}(\mathbf{r})e^{-i\omega_{1}t}+E_{2}(t)u_{2}(\mathbf{r})e^{-i\omega_{2}t}$,
etc., and the atomic 
polarization can be eliminated adiabatically \cite{Hodges,JohnBusch}. 
For class-A lasers, where
the radiative decay rate~$\gamma_{\perp}=\Delta\omega_{a}/2$,
the non-radiative decay rate~$\gamma_{\parallel}$, and the cavity mode
decay rates~$\kappa_j$ are related 
as $\gamma_{\perp}\gg\gamma_{\parallel}\gg\kappa_j$,
the following system of equations 
for the slowly varying envelopes~$E_{j}(t)$ in the two modes 
is obtained \cite{Hodges} ($i,j=1,2,\ i\neq j$),
\begin{widetext}\begin{equation}
\frac{dE_{j}(t)}{dt}=gR_{j}\left[\left(\mathcal{L}_{j}-\frac{\kappa_{j}}{gR_{j}}\right)-\eta\mathcal{L}_{j}\left(\alpha_{jj}^{jj}\mathcal{L}_{j}\left|E_{j}\right|^{2}+\left[\alpha_{ii}^{jj}\mathcal{L}_{i}-\alpha_{ij}^{ji}\mathrm{Re}\left(\chi_{j}\mathcal{M}_{ji}\right)\right]\left|E_{i}\right|^{2}\right)\right]E_{j}(t)+F_{j}(t)
\label{eq:anal_fields}\end{equation}
\end{widetext}
In Eqs.~(\ref{eq:anal_fields}) the terms linear in $E_j(t)$
describe stimulated emission driving and 
are controlled by the light-matter coupling
$g\simeq\sqrt{2\pi\omega_{0}d^{2}/\hbar}$, with $d$ the dipole moment 
of the atomic transition, 
by the pumping rates projected onto the modes $j=1,2$,
$R_{j}=\int_{G}u_{j}^{*}(\mathbf{r})u_{j}(\mathbf{r})R(\mathbf{r})d\mathbf{r}$, and by the cavity mode decay rates $\kappa_j$. 
The coefficients $\mathcal{L}_{j}=\mathrm{Im\,}\beta_{j}^{-1}$, 
with  $\beta_{1,2}=\delta\pm\Delta\omega - i \Delta\omega_{a}/2$,
account for the
different mode-to-gain couplings due to asymmetrical detuning of the 
atomic transition with respect to the resonator lines.
The terms cubic in $E_j(t)$ describe field saturation above the lasing 
threshold, where $\eta=d^{2}/2\gamma_{\parallel}\hbar^{2}$ and 
the overlap integrals
$\alpha_{kl}^{ij}=\int_{G}u_{i}^{*}(\mathbf{r})u_{j}(\mathbf{r})u_{k}^{*}(\mathbf{r})u_{l}(\mathbf{r})d\mathbf{r}$
are taken over the regions $G$ containing the gain medium.
Since $\left|u_{1}(r)\right|^{2} \approx \left|u_{2}(r)\right|^{2}$ we can
assume $\alpha_{jj}^{ii}=\alpha_{ji}^{ij}\equiv\alpha$,
$R_{1}=R_{2}=R$ and $\kappa_{1}=\kappa_{2}=\kappa$.
The cross-saturation terms, with 
$\mathcal{M}_{ij}=\beta_{i}^{-1}+\left(\beta_{j}^{*}\right)^{-1}$ and
$\chi_{1,2}=-i\gamma_{\parallel}/
\left(\pm \Delta\omega -i\gamma_{\parallel}\right)$, 
depend in an assymetrical way on the mode indeces $i,j$.
However, this asymmetry remains small 
 unless $\Delta\omega\ll\Delta\omega_{a}$.

The inhomogeneous terms~$F_{j}(t)$ originate from
the external injection seeding field and from a noise field accounting 
for spontaneous emission \cite{Hodges}. 
For vanishing functions $F_{j}$, Eqs.~\eqref{eq:anal_fields} would
take the form of the standard two-mode competition equations 
\cite{Siegman}
with mode coupling constant~$C$ slightly exceeding unity. This corresponds
to bistable lasing \cite{BabaDisks} and to mode hopping in the
presence of stochastic noise in the system \cite{Bang}. 
If both an external seeding field $\mathcal{E}^{s}(\mathbf{r},t)$ 
and a stochastic noise field $\mathcal{E}^{n}(\mathbf{r},t)$ are present
in the cavity, 
$\mathcal{E}(\mathbf{r},t)= \mathcal{E}^{s}(\mathbf{r},t)+
\mathcal{E}^{n}(\mathbf{r},t)$, one obtains, 
\begin{equation}
\begin{aligned}F_{j}(t)\approx & \frac{\omega_{j}\mathcal{L}_{j}}{\tau}\int_{t-\tau}^{t}dt'e^{i\omega_{j}t'}\int_{G}u_{j}(\mathbf{r})\mathcal{E}(\mathbf{r},t')d\mathbf{r}\\
= & F_{j}^{\mathrm{s}}F(t)+F_{j}^{\mathrm{n}}(t)\ .\end{aligned}
\label{eq:anal_F}\end{equation}
The time integration in Eq.~(\ref{eq:anal_F}) 
is the averaging over a time interval larger than $1/\Delta\omega$.
The function~$F(t)$ is determined by the temporal dependence
of $\mathcal{E}^{s}(\mathbf{r},t)$. 
The coefficients $F_{j}^{\mathrm{s}}$~and~$F_{j}^{\mathrm{n}}(t)$
are determined by the spatial overlap of each mode with the seeding
and noise fields, respectively. We consider the situation when the
seeding prevails over the noise, i.e., $F_{j}^{\mathrm{s}}F(t)\gg F_{j}^{\mathrm{n}}(t)$,
before and during the onset of lasing. After the onset the $E_{j}$ become
so large that the terms~$F_{j}$ have no effect anymore. In this
situation the evolution of the resonator will be determined by
the ratio of $F_{1}^{\mathrm{s}}$~and~$F_{2}^{\mathrm{s}}$. 

\begin{figure}[b]
\includegraphics[width=0.95\columnwidth]{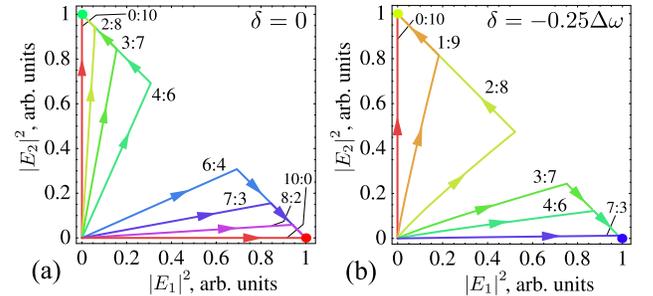}
\caption{(Color online) Cavity phase diagrams for a lasing system governed
by Eq.~\eqref{eq:anal_fields} for $F_j^{\mathrm{s}}F(t)\gg F_j^{\mathrm{n}}(t)$
for (a) symmetric ($\delta=0$) and (b) non-symmetric 
($\delta=-0.25\Delta\omega$) values of the mode frequencies
with respect to the central gain frequency $\omega_a$ 
($\omega_a<\omega_0$, as shown in Fig.~\ref{fig:setup} (b)). 
The dots denote the stable cavity states,
and the curves represent the phase trajectories for their temporal
evolution for different ratios $F_{1}^{\mathrm{s}}:F_{2}^{\mathrm{s}}$
in the direction of the arrows.
\label{fig:phasemap}}
\end{figure}

We analyze the influence of the balance between $F_{1}^{\mathrm{s}}$~and~$F_{2}^{\mathrm{s}}$
on the final lasing state by numerical solution of 
Eq.~(\ref{eq:anal_fields}) and
plotting the phase trajectories of the temporal resonator state evolution  
in the ($\left|E_{1}\right|^{2},\,\left|E_{2}\right|^{2}$)
plane (Fig.~\ref{fig:phasemap}).
The numerical values of the coefficients in Eq.~\eqref{eq:anal_fields}
are calculated for the model PhC structure of Fig.~\ref{fig:setup} a).
As seen in Fig.~\ref{fig:phasemap}, the lasing state first reaches
overall intensity saturation 
($\left|E_{1}\right|^{2}+\left|E_{2}\right|^{2}=E_{s}^{2}$)
and then drifts towards one of the stable fixed points corresponding to
single-mode lasing (either $\left|E_{1}\right|^{2}=E_{s}^{2}$ or
$\left|E_{2}\right|^{2}=E_{s}^{2}$). The drift occurs after the sharp
bend seen in each of the phase trajectories. If the mode coupling 
constant is $C=1$, the drift becomes infinitely slow, and the 
(1,0)--(0,1) line in Fig.~\ref{fig:phasemap} turns into a line of
fixed points. For our case, where $C$
only slightly exceeds unity, the drift happens on a longer time scale
than the initial overall intensity growth, and
the intermode beats decay fast after the lasing onset.
In the case of symmetric detuning of the cavity modes with respect
to the gain frequency ($\delta=0$), single-mode lasing is achieved into that
mode whose spatial overlap with the seeding field, $F_{j}^{\mathrm{s}}$, 
is largest [Fig.~\ref{fig:phasemap} a)].
The asymmetry of the modes with respect to gain ($\delta\neq0$) shifts
the turning point towards one of the modes, but if the seeding is chosen
in a way that the spatial overlaps in Eq.~\eqref{eq:anal_F} result
in $F_{1}^{\mathrm{s}}\gg F_{2}^{\mathrm{s}}$ 
or $F_{1}^{\mathrm{s}}\ll F_{2}^{\mathrm{s}}$,
each of the modes can nonetheless be selected for lasing 
[Fig.~\ref{fig:phasemap} b)].

To refine the predictions of this simple theory of bistable lasing we
have modeled the lasing action in coupled PhC defects with a realistic
injection mechanism (Fig.~\ref{fig:setup})
using the finite-difference time-domain (FDTD) method \cite{lasIEEE,lasRandom}.
The defect modes are located inside the PhC band gap. 
Both defects are filled with an active medium whose population dynamics are
described at each space point 
by the rate equations of a four-level laser with an external pumping rate
$W_p$. In defining the model
and its parameter values we follow in detail Refs.~\cite{lasRandom,lasPauli}.
In particular, the non-radiative transition times of this model
are taken so as to achieve 
population inversion, i.e. $\tau_{32}\simeq\tau_{10}\ll\tau_{21}$, with
$\tau_{31}=\tau_{10}=1\times10^{-13}\,\textrm{s}$,
$\tau_{21}=3\times10^{-10}\,\textrm{s}$, and the
total level population is $N_{\textrm{total}}=10^{24}$
per unit cell \cite{lasPauli}.  The Maxwell equations,
supplemented by the usual equation of motion for the  
polarization density in the medium and by the laser rate equations
\cite{lasIEEE,lasRandom,lasPauli,lasJoannop}, are 
solved for the geometry of Fig.~\ref{fig:setup} a) in TM polarization,
where $\mathbf{E}(\mathbf{r},t)=E_{y}(x,z,t)\, \mathbf{\hat{y}}$. 
The seeding signal is excited by two emitters (linear groups of dipoles)
engineered on the same chip as the PhC and is 
transmitted to the defects through waveguides
in the PhC (see Fig.~\ref{fig:setup}). 
Each of the emitters generates a single short Gaussian pulse with carrier 
frequency~$\omega$ at or near~$\omega_{a}$. The calculations have been
performed using different, fixed values of the half-width duration 
$\sigma_t$ in the range between $\sigma_{t}=5 \times 10^{-14}$ and $10^{-13}$~s.
The relative phase of the fields in these pulses is chosen $0$ or $\pi$.
As expected, such seeding patterns almost exclusively excite
the bonding and antibonding mode, respectively.
Technically, the seeding dipoles are
realized as pointlike oscillating current sources in the Maxwell
equations \cite{Zhukovsky}. 
Similarly, the spontaneous emission \cite{lasRandom,lasSeo,Langevin}
is modeled as an ensemble of point current sources, randomly placed in space,
with temporally $\delta$-correlated Langevin noise \cite{Langevin}.      
For the FDTD computations the computational domain
$13a\times13a$ with perfectly matched layer (PML) boundary conditions
was discretized with an $a/16$ mesh and a time step of 
$dt=6\times10^{-17}\,\textrm{s}$ required by numerical stability,
see \cite{AVLCharact} for details.

\begin{figure}
\includegraphics[clip,width=0.95\columnwidth]{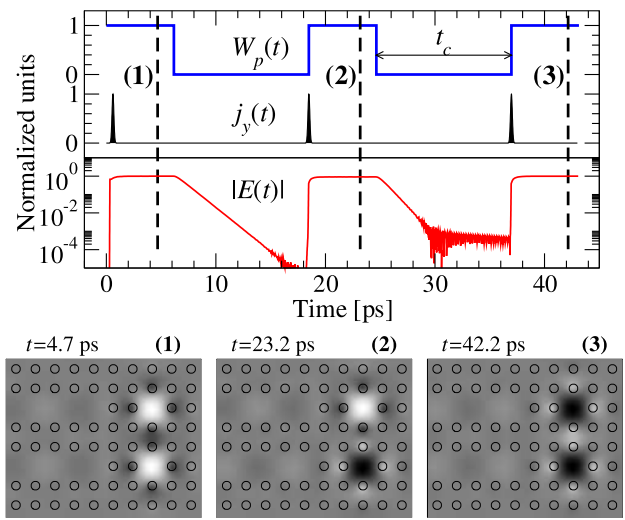}
\caption{(Color online) Illustration of mode-to-mode switching. \emph{Top}:
time dependence of the normalized pumping power~$W_{p}$, seeding
signal $j_{y}$, and cavity field~$E_{y}$. \emph{Bottom}: spatial
distribution of the electric field amplitude $E_{y}(x,z)$ 
(grey scale) in the steady-state
lasing regime after initial seeding (1) and each re-seeding (2,~3).
The time instants (1)--(3) are marked in the top panel by dashed lines.
This calculation was done for a typical pumping rate 
$W_{p}=1\times10^{13}\,\textnormal{s}^{-1}$ and a seeding pulse 
duration  $\sigma_{t}=1.2\times10^{-13}$ s. 
\label{fig:switching}}
\end{figure}

Neglecting noise at first, we find that, if the laser
amplification line spectrally covers both resonant modes and provides
a comparable effective gain for each of them 
(i.e. $\left|\delta\right|<\Delta\omega$), a seeding signal of the type
described above can individually select any of the two modes when applied (with
arbitrary strength) during the onset of lasing, i.e. during the 
exponential growth after switching on the pumping. 
The steady-state lasing is then nearly single-mode, and the dominant mode
is the one whose symmetry matches that of the seeding signal 
(Fig.~\ref{fig:switching}). This is in agreement with the semianalytical
theory described above. The electric field maps in 
Fig.~\ref{fig:switching} (bottom) show that the spatial field distribitions 
of the two lasing modes remain nearly unaffected by the side-coupled 
seeding waveguides. In the present setup the laser output is 
primarily delivered through the seeding waveguides.
A detailed study of the microlaser radiation 
characteristics will be given elsewhere.  
Moreover, we have shown that successive switchings from one mode to another
are possible. To achieve this, the pump is first turned off
to allow the currently lasing mode to decay. After a certain cool-down
time~$\tau_{c}$, the pumping is turned back on, and the cavity is re-seeded
for the other mode in the same way as the initial seeding occurred.
Fig.~\ref{fig:switching} demonstrates such a switching sequence from
the bonding to the antibonding mode and back. The pump was kept
on for a time needed to achieve quasi-continuous lasing, as confirmed
by the temporal dependence of the cavity field. Looking at the spatial
field distribution in this regime 
[panels (1)--(3) in Fig.~\ref{fig:switching}],
we make sure that the mode switching occurs in the desired order.

\begin{figure}
\includegraphics[clip,angle=-90,width=0.9\columnwidth]{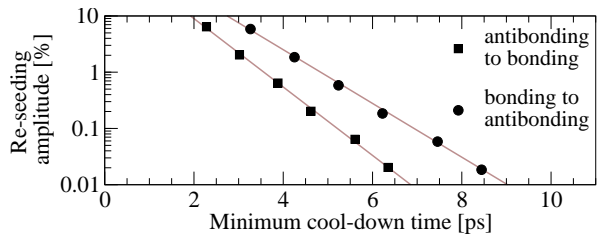}
\caption{(Color online) 
The minimum switching time $\tau_{c,min}$ versus the
re-seeding amplitude, normalized to the
saturated lasing mode amplitude. The seeding pulse duration is 
$\sigma_t=0.12\,\mathrm{ps}$. The cavity mode decay time is 
$\tau=0.7\,\mathrm{ps}$. The longer $\tau_{c,min}$
for switching from the bonding to the antibonding mode is due to the
non-zero gain detuning $\delta$ towards the bonding mode frequency
[see Fig.~\ref{fig:setup} b)].
\label{fig:tau_cmin}}
\end{figure}

The switching time $\tau_s$ between two lasing modes 
is primarily controlled by the minimum cool-down time $\tau_{c,min}$, i.e. 
by the relatively slow mode decay time, $\tau=1/\kappa$, and hence 
by the cavity Q-factor $Q=\omega_j/\kappa$, not by the
fast lasing onset after re-seeding, $\tau_s\approx \tau_{c,min}$. 
Since the mode decays exponentially with time from its steady-state
lasing amplitude $A_0$ and 
the re-seeding signal must be strong enough to override the
residue of the decaying initial mode,
$\tau_{c,min}$ decreases logarithmically with increasing re-seeding 
pulse power (i.e. with its time-integrated intensity) or, for a 
Gaussian pulse, with the seeding amplitude $S$, 
$\tau_{c,min}= - (c_j/\kappa)\,\ln (S/A_0)$. Here, $c_j$ is a dimensionless
factor describing the coupling of the cavity mode $j=1,2$ to the 
laser transition. The corresponding numerical FDTD results are shown
in Fig.~\ref{fig:tau_cmin}. For realistic parameter values 
(Fig.~\ref{fig:tau_cmin}) switching times of a
few tens of picoseconds can be realized for re-seeding amplitudes $S$ as
low as 0.001 per cent of the lasing mode amplitude $A_0$, two to three
orders of magnitude faster than in previous devices \cite{Lavrova}.

When noise is present in the cavity, the FDTD calculations show that
controlled switching is preserved as long as the integrated seeding power
exceedes the noise power integrated over the lasing onset time.
Otherwise, noise begins to dominate the lasing spectrum
formation. In this case, the mode whose spatial overlap with the noise field at
the onset of lasing is larger wins the competition. The same happens
when the seeding pulse does not match the onset of lasing in time
and its field residue has smaller amplitude in comparison to the noise
field. These numerical observations are fully consistent with the analysis
of the competition equations \eqref{eq:anal_fields} with the functions
$F_{j}(t)$ given by \eqref{eq:anal_F}. The switching is effective
only if the influence of the seeding prevails in the resonator at
the period of time when laser radiation starts to build up\emph{.}

To summarize, the concept of \emph{switchable} (rather than tunable)
lasing in microstructures has been introduced. Instead of externally
changing the parameters of a single-mode cavity, an inherently multimode
cavity is used, and one of the modes is deliberately made to be dominant
for lasing by means of injection seeding. This offers the possibility
of all-optical frequency selection and switching in microlasers
with particularly low switching times. As an example,
we have investigated the mode switching in a system of two coupled
defects in a 2D PhC lattice. For realistically chosen parameters
a mode-to-mode switching on the picosecond scale 
has been numerically demonstrated.
The results are consistent with a qualitative semianalytical model. It
shows that a resonator supporting modes with similar spatial intensity
profile tends towards bistability, which is the underlying physical
mechanism of switchable lasing. 
The proposed concept is not limited to the model considered,
but is expected to work in any resonator featuring bi- or multistability.
Any coupled cavity based system would be a good candidate for the
effects predicted. For instance, we have observed four-mode switching
in coupled nanopillar waveguides of both periodic and non-periodic
longitudinal geometry \cite{Zhukovsky}. 
\begin{acknowledgments}
This work was supported in part by DFG through SPP 1113 and 
FG 557 (SVZ, DNC, and JK) and by the EU Commission FP6 via  
project NewTon, NMP4-CT-2005-017160 (AVL). 

\end{acknowledgments}

\end{document}